\documentclass{emulateapj}

\shorttitle{SiO thermal line variability toward L1448-mm}
\shortauthors{I. Jim\'enez-Serra et al.}

\begin{document}

\title{Variability of the SiO thermal line emission toward the young 
L1448-mm outflow}

\author{I. Jim\'{e}nez-Serra\altaffilmark{1}, J. Mart\'{\i}n-Pintado\altaffilmark{2}, J.M. Winters\altaffilmark{3}, A. Rodr\'{\i}guez-Franco\altaffilmark{2,4}, and P. Caselli\altaffilmark{5}}

\altaffiltext{1}{Harvard-Smithsonian Center for Astrophysics, 
60 Garden St., Cambridge, MA 02138, USA; ijimenez-serra@cfa.harvard.edu}

\altaffiltext{2}{Centro de Astrobiolog\'{\i}a (CSIC/INTA),
Ctra. de Torrej\'on a Ajalvir km 4,
E-28850 Torrej\'on de Ardoz, Madrid, Spain; 
martinpmj@inta.es,
arturo@damir.iem.csic.es}

\altaffiltext{3}{Institut de Radio Astronomie Millim\'etrique, 300 Rue 
de la Piscine, F-38406 St. Martin d'H\`eres, France; winters@iram.fr}

\altaffiltext{4}{Escuela Universitaria de \'Optica,  
Departamento de Matem\'atica Aplicada (Biomatem\'atica),
Universidad Complutense de Madrid,
Avda. Arcos de Jal\'on s/n, E-28037 Madrid, Spain}

\altaffiltext{5}{School of Physics \& Astronomy, E.C. Stoner Building,
The University of Leeds, Leeds, LS2 9JT, UK; P.Caselli@leeds.ac.uk}

\begin{abstract}

The detection of narrow SiO thermal emission toward young outflows has been proposed to be a signature of the magnetic precursor of C-shocks. Recent modeling of the SiO emission across C-shocks predicts variations in the SiO line intensity and line shape at the precursor and intermediate-velocity regimes in only few years. We present high-angular resolution (3.8$''$ $\times$3.3$''$) images of the thermal SiO $J$=2$\rightarrow$1 emission toward the L1448-mm outflow in two epochs (November 2004-February 2005, 
March-April 2009). Several SiO condensations have appeared at intermediate velocities (20-50$\,$km$\,$s$^{-1}$) toward the red-shifted lobe of the outflow since 2005. Toward one of the condensations (clump D), systematic differences of the dirty beams between 2005 and 2009 could be responsible for the SiO variability. At higher velocities (50-80$\,$km$\,$s$^{-1}$), SiO could also have experienced changes in its intensity. We propose that the SiO variability toward L1448-mm is due to a real SiO enhancement by young C-shocks at the internal working surface between the jet and the ambient gas. For the precursor regime (5.2-9.2$\,$km$\,$s$^{-1}$), several narrow and faint SiO components are detected. Narrow SiO tends to be compact, transient and shows elongated (bow-shock) morphologies perpendicular to the jet. We speculate that these features are associated with the precursor of C-shocks appearing at the interface of the new SiO components seen at intermediate velocities. 

\end{abstract}

\keywords{star: formation --- physical data and processes: shock waves ---
ISM: jets and outflows --- ISM: individual (L1448)}

\section{Introduction}

Variability is a common phenomena in the evolution of very young 
stars. \citet{joy45} very early found that T Tauri stars have an 
irregular behavior in their optical emission. More recently, 
simultaneous X-ray and optical/near-IR variability studies 
have shown that a high fraction of pre-main-sequence
stars are variable in X-ray and optical/near-IR
emission \citep{sta06,for07}. 
At radio wavelengths, young stellar objects are known to show 
variability in their radio continuum emission 
in time-scales of few years \citep[][]{cur06,pech10}. 
In the case of molecular line emission, only {\it non-thermal} maser 
emission has been found to be variable
\citep[see][]{reid81,eli92}. However, variability in the morphology, line 
intensity and line profiles of {\it thermal} molecular emission 
still remains to be reported.

Broad SiO thermal line emission
is known to be a good indicator of shocked gas in molecular outflows 
\citep[][]{mar92}. The SiO line profiles have typical linewidths of 
tens of km$\,$s$^{-1}$ and appear shifted with respect to the 
ambient cloud velocity. In addition to this broad emission, 
\citet{lef98} and \citet{jim04} reported the detection of very narrow 
SiO thermal lines (linewidths of $\leq$1$\,$km$\,$s$^{-1}$) centered
at almost ambient velocities toward the young NGC1333 and L1448-mm 
outflows. Toward L1448-mm, \citet{jim04}
proposed that narrow SiO could arise from the magnetic precursor region within very young C-shocks \citep{jim04}. 

At the early stages of the C-shock propagation, the magnetic precursor 
of C-shocks accelerates, compresses and heats the ion fluid before the 
neutral one \citep[][]{dra80}. The subsequent ion-neutral velocity
drift generates enough friction to sputter dust grains 
\citep{cas97,gus08}, progressively enhancing the gas-phase abundances 
of SiO \citep{jim08}. \citet{jim09} have claimed that the SiO line 
profiles in young outflows should then evolve from very narrow lines 
arising from the magnetic precursor region, to broader SiO emission 
generated in the intermediate- and high-velocity postshock gas. 
From this, variability of the narrow SiO lines and of the 
intermediate-velocity SiO emission should appear in only few years \citep{jim09}. 

We present interferometric images of the SiO $J$=2$\rightarrow$1 thermal line 
emission toward the L1448-mm outflow observed with the PdBI at two different 
epochs (2005 and 2009)\footnote{Based on observations carried 
out with the IRAM Plateau de Bure Interferometer. IRAM is supported by 
INSU/CNRS (France), MPG (Germany) and IGN (Spain)}.
These images show that several SiO condensations (named A, B, C, D and E) have recently appeared at intermediate velocities along the direction of the red-shifted lobe of the outflow since 2005. The increase in the SiO flux toward condensations A, B and C are likely real and produced by an enhancement in the gas-phase abundance of SiO by the recent interaction of young C-shocks with the ambient gas. The variable nature of the SiO clumps D and E, however, 
remains unclear (Section$\,$\ref{mod}). Our results also reveal the detection in 2009 of several new narrow SiO components that tend to be found at the interface between the
SiO condensations A, B, C, D and E reported at intermediate velocities 
toward the red-shifted lobe of the outflow (Section$\,$\ref{pre}). We speculate
that the recently detected narrow SiO components are possibly associated
with the magnetic precursor of C-shocks (Section$\,$\ref{origin}). This is the first time that variability of molecular thermal emission is ever reported. 

\section{Observations}
\label{obs}

PdBI observations of the thermal 
SiO $v$=0 $J$=2$\rightarrow$1 line emission (at 86.846$\,$GHz) 
toward the L1448-mm outflow, were carried out in the 
B and C configurations in November 2004-February 2005 
(for $\sim$15$\,$hrs of on-source time) and in the C configuration 
in March-April 2009 ($\sim$9$\,$hrs). The phase center of the images was set at 
$\alpha(J2000)$=03$^{h}$25$^{m}$38.8$^s$, 
$\delta(J2000)$=+30$^{\circ}$44$'$05$''$. The correlator setup 
included two different spectral resolution units (with bandwidths of 
20$\,$MHz and 80$\,$MHz), which provided spectral resolutions of 
39$\,$kHz and 312$\,$kHz, respectively (i.e. velocity resolutions of 
0.14 and 1.1$\,$km$\,$s$^{-1}$ at $\sim$87$\,$GHz). 
We used 3C84 (9.5$\,$Jy) as bandpass calibrator,
and MWC349 (1.2$\,$Jy) as flux calibrator. 0333+321 (2$\,$Jy) and 0234+285 
(2$\,$Jy) were used as gain calibrators. Pointing and focus 
of the antennas was checked every 65$\,$min and re-adjusted 
if necessary. Pointing errors were typically $\sim$2$"$ 
(i.e. less than 4$\%$ the field-of-view, FOV, of 58$"$ at 
$\sim$87$\,$GHz) in the 2005 observations, and $\sim$2-4$"$ 
($\sim$4-8$\%$ the FOV) in the 2009 tracks. Focus corrections 
were $\leq$0.1$\,$mm in the 2005 observations, and 
$\sim$0.05-0.15$\,$mm in the 2009 tracks. Data calibration was done with the GILDAS package (CLIC)\footnote{See http://www.iram.fr/IRAMFR/GILDAS.}. We used the $"$Data Quality Assessment$"$ option available within this software, to make sure that 
only data with a phase noise smaller than 40$^\circ$ were retained for constructing the final maps.

\section{UV data analysis and imaging}
\label{uv}

Construction of the UV tables, data imaging, continuum subtraction and cleaning were also carried out within GILDAS by using CLIC and MAPPING. Independent UV tables were generated 
for the 20$\,$MHz and 80$\,$MHz correlator units used in the observations. 
The UV data were resampled in velocity by imposing the same number of 
channels (460 and 230 for the 20$\,$MHz and 80$\,$MHz units, respectively), 
the same position for the central channel (230 for the 20$\,$MHz unit, and 115 for the 80$\,$MHz unit), LSR velocity of the source 
($-$5$\,$km$\,$s$^{-1}$ in all cases) and channel spacing 
(39$\,$kHz or 0.14$\,$km$\,$s$^{-1}$ for the 20$\,$MHz unit, and 312$\,$kHz or 
1.1$\,$km$\,$s$^{-1}$ for the 80$\,$MHz unit). 

\subsection{The 80$\,$MHz low spectral resolution data}
\label{80MHz}

The SiO emission toward the L1448-mm outflow is known to be detected to 
velocities up to $\pm$80$\,$km$\,$s$^{-1}$ \citep[][]{mar92,jim04,jim05}. 
The velocity coverage provided by the 80$\,$MHz correlator units was 
$\sim$276$\,$km$\,$s$^{-1}$, covering the entire SiO line profile at low spectral 
resolution (312$\,$kHz or 1.1$\,$km$\,$s$^{-1}$). The morphology of the SiO $J$=2$\rightarrow$1 emission 
at intermediate (from 20 to 50$\,$km$\,$s$^{-1}$) and high velocities (from 50 to 
80$\,$km$\,$s$^{-1}$) is extended \citep{gui92,gir01} 
and therefore sensitive to variations of the UV plane coverage.
In particular, since the configuration of the PdBI array in 2009 was more compact than in 2005, 
the 2009 observations could have been more sensitive to extended emission
introducing a low-level contribution that could have mimicked a time-variation 
in the flux and morphology of the SiO emission. To avoid this potential problem, we have homogenized the 2005 and 2009 datasets in the UV plane by selecting the {\it common} 
visibilities to both epochs contained within UV cells of 10$\,$m-size. 
We created a MAPPING procedure that sets the weight of the 
{\it non-overlapping} visibilities to negative values so that these 
visibilities were flagged out and not considered in the subsequent steps 
for imaging. As shown below, this has the disadvantage 
to decrease the sensitivity of the original observations, but guarantees that any 
possible variation detected in the SiO $J$=2$\rightarrow$1 line profile 
toward L1448-mm is not affected by differential missing flux due to different UV sampling in the two epochs. 

The continuum UV tables were generated by considering the same line-free frequency ranges in both datasets. After selecting the common visibilities in both continuum UV tables, we subtracted the continuum UV data from the line+continuum UV tables for the 2005 and 2009 epochs by using the UVSUBTRACT task within MAPPING.      

The dirty images were created using the UVMAP task within MAPPING. 
The size of the dirty maps was 256$\,$pixels$\times$256$\,$pixels, with pixel size 
of 0.86$"$$\times$0.86$"$. We used robust weighting to supress the 
contribution from the secondary lobes of the beams. The resulting 
synthesized beams were very similar (see also Section$\,$\ref{mod}). For the 
line dirty images, the beam sizes were of 3.80$"$$\times$3.35$"$, P.A.=179$^\circ$ for the 2005 images, and 3.87$"$$\times$3.31$"$, P.A.=157$^\circ$ for the 2009 maps. 
For the continuum datacubes, the synthesized beams were 
3.48$"$$\times$3.33$"$, P.A.= 28$^\circ$ for the 2005 epoch, and 
3.44$"$$\times$3.37$"$, P.A.=129$^\circ$ for the 2009 images. 

We finally cleaned the dirty images by using the Hogbom 
method within MAPPING. We considered for all images a circular support 
of 60$"$-diameter (comparable to the FOV). 
The stopping criteria for the CLEAN task was met for every channel 
when the maximum amplitude of the absolute value of the residual image 
was lower than 2.5\% the peak intensity of the dirty map. We note that 
this criterion is set up by default within the MAPPING software. 
The final analysis of the cleaned images was carried out
with the new MADCUBAJ software\footnote{See http://www.damir.iem.csic.es/mediawiki-1.12.0}.

An automatic data analysis procedure was applied to the UV data 
to calculate the visibility noise for every single channel in the 
UV tables. In Table$\,$\ref{tab1}, we show the median 
values obtained from the histograms of the visibility 
noise per channel for the 80$\,$MHz unit tables. The noise is almost 
a factor of 2 larger in the 2005 UV data ($\sim$23$\,$mJy/beam per 312$\,$kHz-channel) 
with respect to the 2009 dataset ($\sim$13$\,$mJy/beam per 312$\,$kHz-channel). 
Hereafter, we will use the thermal noise of the 
2005 data as a reference to search for differences in the 
morphology and intensity of the SiO line emission at intermediate and high
velocities between both 
epochs. We note that the thermal noise of the original 
2005 and 2009 UV tables (i.e. considering the common and the non-overlapping UV data) ranged between 5 and 7$\,$mJy/beam per 312$\,$kHz-channel, 
i.e. factors of 2-3 better than those obtained in the common
UV data tables.

\citet{mas86} proposed another technique to measure 
very small changes in the structure of very bright radiocontinuum 
sources such as the planetary nebula NGC7027. The cross-calibration 
method is based on the use of the closure phases and amplitudes 
derived from the data from epoch I, to minimize the effects of 
calibration errors on the final subtracted map. 
This method, however, assumes that the source is bright enough to do self-calibration. The 3$\,$mm continuum flux measured toward the L1448-mm source is only $\sim$10$\,$mJy, insufficient to apply this technique (see Section$\,$\ref{mod}).


\begin{deluxetable*}{lcccccl}
\tablecaption{Median values of the visibility noise per channel for all derived UV tables\label{tab1}}
\tablewidth{0pt}
\tablehead{ 
& & & \multicolumn{3}{c}{Visibility Noise} \\ \cline{4-6}
\colhead{Bandwidth} & \colhead{Chan. Res.} & \colhead{Epoch} & \colhead{Thermal} & \colhead{Real} & \colhead{Imaginary} & \colhead{}}
\startdata
80$\,$MHz & 312$\,$kHz & 2005 & 22.8 & 23.1 & 23.2 & mJy/beam \\
80$\,$MHz & 312$\,$kHz & 2009 & 12.6 & 14.4 & 14.4 & mJy/beam \\ \hline
20$\,$MHz & 39$\,$kHz & 2005 & 21.8 & 22.0 & 21.7 & mJy/beam \\
20$\,$MHz & 39$\,$kHz & 2009 & 18.7 & 21.1 & 21.1 & mJy/beam
\enddata
\end{deluxetable*}

\subsection{The 20$\,$MHz high spectral resolution data}
\label{20MHz}

With the high-spectral resolution 20$\,$MHz units, we imaged the 
very narrow SiO $J$=2$\rightarrow$1 line emission found toward the L1448-mm
outflow at velocities close to the ambient cloud velocity \citep{jim04}. 
This narrow emission is expected to be faint and therefore, the data imaging 
procedure followed in Section$\,$\ref{80MHz} cannot be applied to the 20$\,$MHz 
data because of the poor sensitivity achieved in the final images. We believe that
differences in the UV coverage between both epochs do not represent any
problem for our variability studies of the narrow SiO lines toward L1448-mm, because
this emission is found to be compact and to arise from practically unresolved 
condensations associated with the red-shifted lobe of the outflow 
(see Section$\,$\ref{pre}). In any case, we only used the C configuration 2005 UV 
data so that the UV coverage in the 2005 images was kept as close as possible to that 
obtained in the 2009 observations (also carried out in C configuration). 

The 2005 and 2009 datasets were subsequently homogenized by weighting the 2005 UV-data 
(beam of 3.52$''$$\times$2.22$''$, P.A.=35$^{\circ}$) with a Gaussian taper 
of 80$\,$m$\times$100$\,$m and at a P.A.=65$^{\circ}$. The resulting 
2005-beam was 5.20$''$$\times$3.65$''$, P.A.=21$^{\circ}$, which 
matched the 2009-beam of 5.18$''$$\times$3.69$''$, 
P.A.=57$^{\circ}$. The map size and pixel size used to construct the dirty images
were the same as those of the 80$\,$MHz data images. In this case, however, we used
natural weighting instead of robust weighting to increase the final 
signal-to-noise ratio. The dirty images were cleaned in a similar fashion as 
the 80$\,$MHz maps. The derived visibility 
noise per channel for the 20$\,$MHz UV datasets is shown in Table$\,$\ref{tab1}. 
Like in the 80$\,$MHz data, the noise in the 2005 observations was 
higher (21.8$\,$mJy/beam per 39$\,$kHz-channel) than in 2009 (18.7$\,$mJy/beam per 39$\,$kHz-channel). We will then use the 2005 thermal noise level as the reference to search for variability in the narrow SiO lines toward L1448-mm. 

We note that, although the noise in the high-spectral resolution 20$\,$MHz data is expected to be significantly worse than in the 80$\,$MHz UV data, the 20$\,$MHz data considers all visibilities available from the 2005 and 2009 observations, while the low-spectral resolution 80$\,$MHz UV files have been constructed from only UV data common to both epochs.

\section{Results}
\label{res}

\subsection{Variability of the thermal SiO emission at the intermediate- and high-velocity shock regimes}
\label{mod}

\begin{figure*}
\begin{center}
\includegraphics[angle=270,scale=0.75]{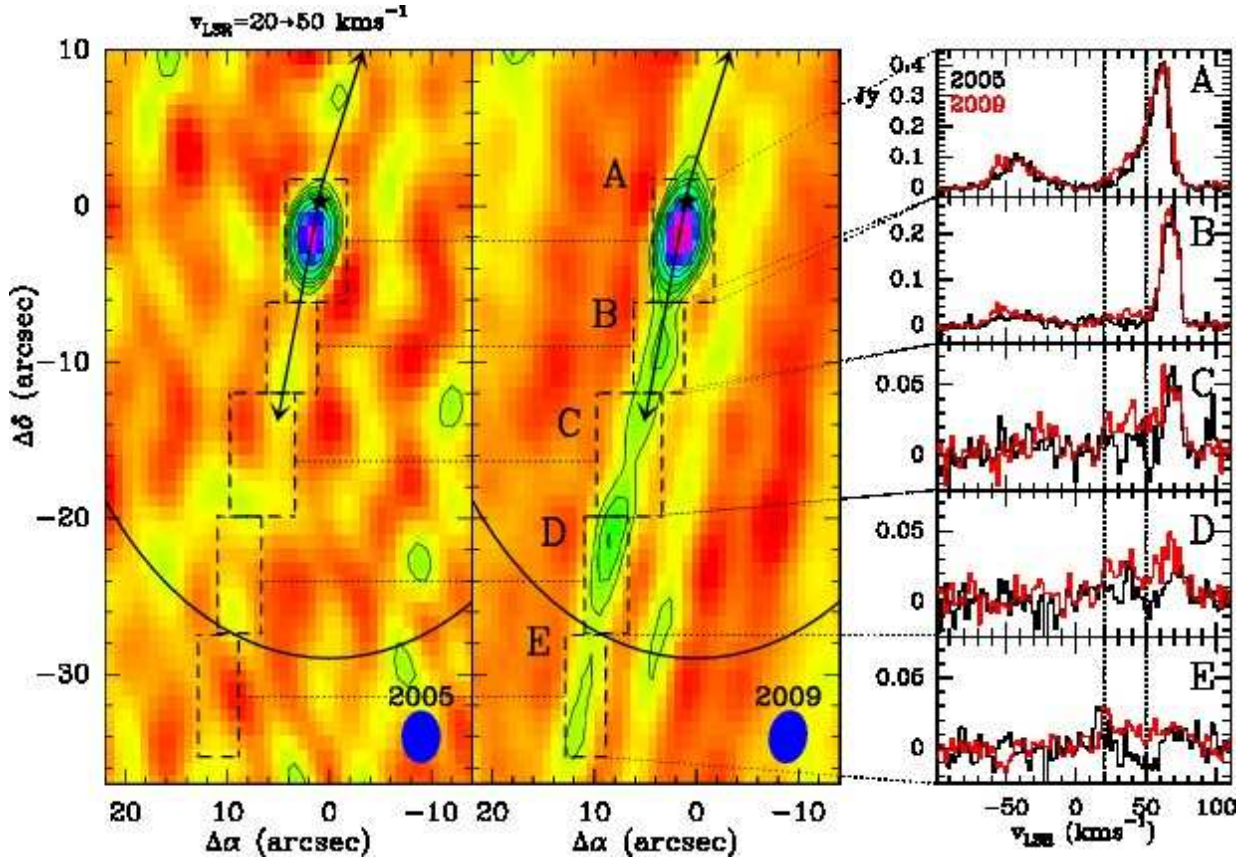}
\caption{{\it Left and middle panels:} Averaged intensity images 
of the SiO $J$=2$\rightarrow$1 emission from 20 to 
50$\,$km$\,$s$^{-1}$ observed toward L1448-mm in 2005 and 2009. 
Contour levels in both maps are 8.8 (2$\sigma$), 
13.2, 17.6, 22, 30.8 and 39.6$\,$mJy/beam. 
The 1$\sigma$ level (4.4$\,$mJy/beam) is calculated from the thermal 
noise measured in the 2005 UV dataset (Section$\,$\ref{res} and Table$\,$\ref{tab1}). Negative contours correspond to the $-$3$\sigma$ 
noise level (dotted contours). Arrows show the 
direction of the SiO high-velocity jet with velocities from 
50 to 80$\,$km$\,$s$^{-1}$. 
Filled star gives the location of the L1448-mm source \citep{mau10}. 
Solid ellipse shows the primary beam of our PdBI observations. 
Dashed boxes (A through E) delineate the 
regions where the 2005 and 2009 SiO spectra have been averaged. Beams 
are shown at the lower right corners and are 3.80$"$$\times$3.35$"$, 
P.A.=179$^\circ$ for the 2005 images, and 3.87$"$$\times$3.31$"$, 
P.A.=157$^\circ$ for the 2009 maps. 
{\it Right panels:} SiO spectra averaged within condensations
A, B, C, D and E for the 2005 (black histograms) and 2009 
(red histograms) epochs. The units of the spatially averaged SiO spectra 
are given in units of Jy.}
\label{f1}
\end{center}
\end{figure*}

In Figure$\,$\ref{f1}, we present the averaged intensity images of
the SiO $J$=2$\rightarrow$1 emission from 20 to 50$\,$km$\,$s$^{-1}$ 
observed in 2005 (left panel) and 2009 (central panel) toward the 
red-shifted lobe of the L1448-mm outflow. For our analysis, we use averaged intensity images instead of integrated intensity maps because it will allow a simpler comparison of the data with the response of the interferometer (beam) for both epochs (see below).
Contour levels are the same in both averaged intensity maps and represent the 2, 3, 4, 5, 7 and 9$\sigma$ levels with 1$\sigma$=4.4$\,$mJy/beam. 
This rms level has been estimated from the thermal noise 
measured in 2005 (22.8$\,$mJy/beam per channel; Table$\,$\ref{tab1}) divided by $\sqrt{N}$,
where $N$ is the number of channels. For the velocity range from 
20 to 50$\,$km$\,$s$^{-1}$, the number of channels is $\sim$27. 
As mentioned in Section$\,$\ref{uv}, we use the 2005 thermal noise
as a reference, since this is the largest r.m.s. noise measured between both 
epochs (Table$\,$\ref{tab1}).    

Figure$\,$\ref{f1} shows that the SiO line emission at intermediate velocities has experienced significant changes in 
morphology and intensity since 2005. The SiO emission
in the 2005 image is clearly restricted to one condensation 
close to the L1448-mm protostar at offset (2,-2), emission peak A, which is coincident with the molecular {\it bullet} R1 \citep{gui92,dut97,gir01}. This emission is also seen in the higher-excitation transition SiO $J$=8$\rightarrow$7 imaged by \citet{hir10}. In the 2009 map, however, SiO $J$=2$\rightarrow$1 is also detected toward three other fainter clumps (seen at 
the 2-4$\,$$\sigma$ level)  along the direction of the high-velocity jet. These clumps are located at offsets (3,-9) for B, (6,-16) for C, and (9,-21) for D, and do not seem to have any counterpart in previous observations of the SiO $J$=2$\rightarrow$1 emission carried out with the IRAM PdBI with comparable sensitivity \citep[see Figures$\,$2, 3 and 4 in][]{dut97}. Indeed, the position-velocity diagrams of SiO obtained along the outflow \citep{dut97} show that the region where the new B, C and D clumps are found (from 10$"$ to 30$"$ south the L1448-mm central source) was devoid of SiO emission in 1993, supporting the idea of an increase in the SiO intensity since 2005. Outside the primary beam, another very faint condensation, clump E at offset (11, -34), is also found along the jet probably belonging to the same global structure of the new SiO clumps. This clump could be associated with the molecular bullet R2 already detected in the SiO maps of \citet{dut97}. We note that the condensations B, C, D and E are unresolved in the direction perpendicular to the high-velocity jet with measured sizes of $\leq$4$"$. The derived deconvolved sizes are of $\leq$1.8$"$. From all this, we propose that the new SiO condensations detected at intermediate velocities are due to an increase in the SiO gas-phase abundance generated by the recent interaction of young C-shocks with the ambient gas.

\begin{figure*}
\begin{center}
\includegraphics[angle=270,scale=0.7]{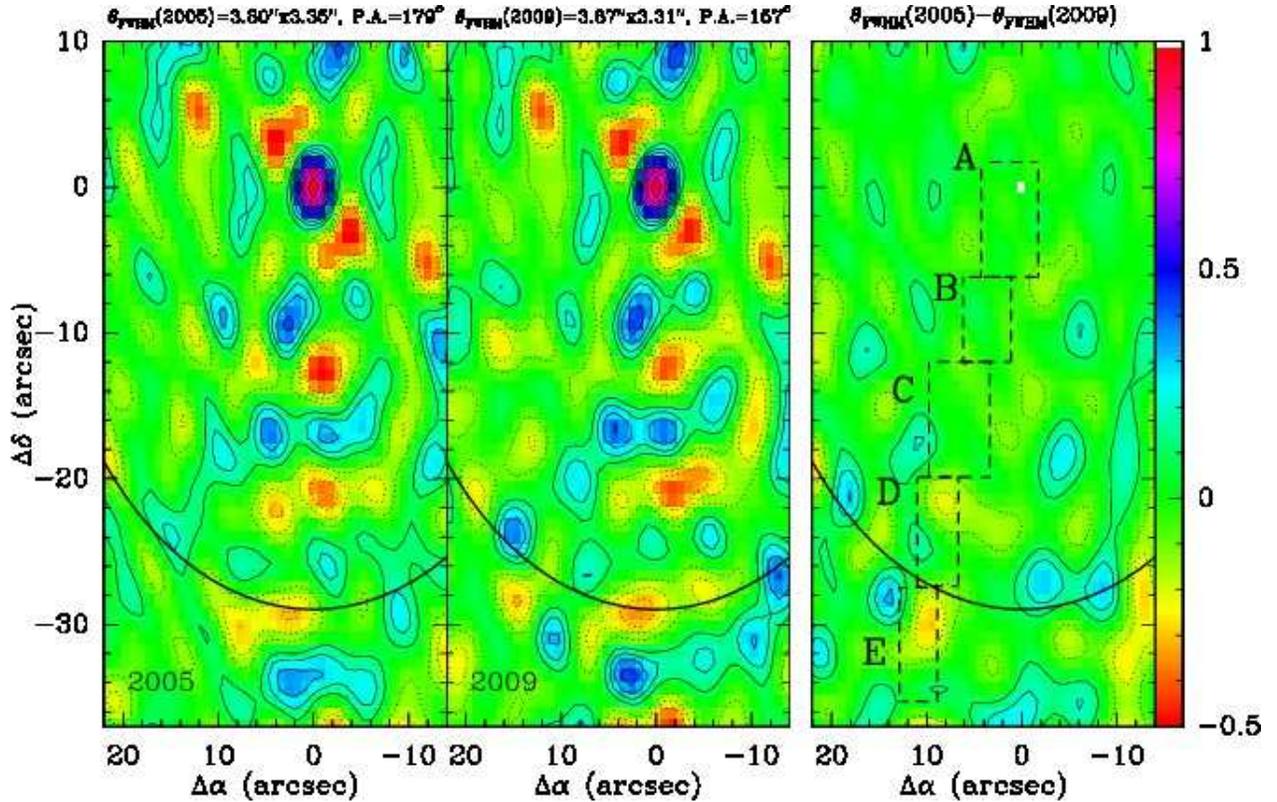}
\caption{Comparison of the beams obtained in the 2005 and 2009 
observations with the PdBI (left and middle panels), and subtraction 
of the 2009 beam pattern from the 2005 beam (right panel). First (solid) 
contour and step level correspond to 10\% the peak value of the beam. 
Negative (dotted) contours are -10\%, -20\% and -30\% this maximum value.  
Color scale is shown on the right panel, and beam sizes are indicated at the upper part of the left and middle panels. Solid ellipse shows the primary beam (FOV$\sim$58$''$). Dashed boxes (right panel) delineate 
the regions where variability of the SiO emission has been detected (see Figure$\,$\ref{f1}). }
\label{f2}
\end{center}
\end{figure*}

To show that the changes in the morphology and intensity of the SiO emission found in the 2009 images are not a consequence of the non-identical UV coverage of the 2005 and 2009 observations, in the left and middle panels of Figure$\,$\ref{f2} we present the dirty beams of these observations for the region reported in 
Figure$\,$\ref{f1}. In the right panel of Figure$\,$\ref{f2}, we also show the subtraction of the two dirty beams, $\theta_{FWHM}$(2005)-$\theta_{FWHM}$(2009). 
Dashed boxes show the regions where the new SiO condensations have been detected (see Figure$\,$\ref{f1}). 
From Figure$\,$\ref{f2}, it is clear that both beams are very similar
with intensity differences of less than 4.5$\,$mJy/beam (i.e. 0.045 after being normalized to the beam peak flux of 100$\,$mJy/beam) for condensations A, B and C. 
This intensity difference is within the r.m.s. of our observations 
(4.4$\,$mJy/beam), which implies that the SiO intensity changes observed toward A, B, and C are likely due to time variability.

Toward condensation D, the SiO emission reaches the 4$\sigma$ detection level (i.e. 17.6$\,$mJy/beam). However, the absolute value of the intensity differences between the 2005 and 2009 beams gets as large as 23$\,$mJy/beam, making the variable nature of the SiO emission toward this condensation uncertain. For clump E, the intensity differences between the beams reach 27$\,$mJy/beam in absolute value toward the north-west region of box E, where no variation in the SiO emission has been detected. In 
the south-east region of this condensation, the beam differences are less 
than 3$\,$mJy/beam implying that, although faint (intensity at the 2$\sigma$ level, i.e. 8.8$\,$mJy/beam), the new SiO clump seen at 
the border of the primary beam could have also appeared recently.

The increase of the SiO emission flux at intermediate velocities 
in 2009 is also evident from the right panels of Figure$\,$\ref{f1}, 
where we compare the SiO spectra averaged within condensations 
A, B, C, D and E for the 2005 and 2009 observations, and smoothed to a velocity
resolution of 2.2$\,$km$\,$s$^{-1}$. The intensity of
the spatially averaged SiO spectra in Figure$\,$\ref{f1} are given in units 
of Jy. In Table$\,$\ref{tab2} (Columns 3 to 5), we report the SiO integrated intensity flux 
measured in both epochs toward A, B, C, D and E for the 20-50$\,$km$\,$s$^{-1}$ velocity range. In Table$\,$\ref{tab2}, we also include the SiO fluxes derived for the velocity
ranges from -10 to 20$\,$km$\,$s$^{-1}$ and from 50 to 80$\,$km$\,$s$^{-1}$. 
The errors in the integrated intensity flux
are calculated as $\sigma_{Area}$=$\sigma$$\sqrt{\delta v \Delta v}$, where $\sigma$ is the rms noise level
in the SiO spectra per channel (0.008$\,$Jy and 0.006$\,$Jy for the 2005 and 2009 SiO spectra, respectively), $\delta v$ is the velocity resolution (2.2$\,$km$\,$s$^{-1}$), and $\Delta v$ is the total width considered in every velocity range (30$\,$km$\,$s$^{-1}$). 
In Columns 6 to 8 of Table$\,$\ref{tab2}, we show the ratios between the 2005 and 2009 SiO fluxes, $F$(2009)/$F$(2005), and in Columns 9 to 11, the absolute differences between these fluxes $|F$(2009)-$F$(2005)$|$. The 
uncertainties of these quantities have been estimated by propagating errors.

\begin{figure*}
\begin{center}
\includegraphics[angle=270,scale=0.8]{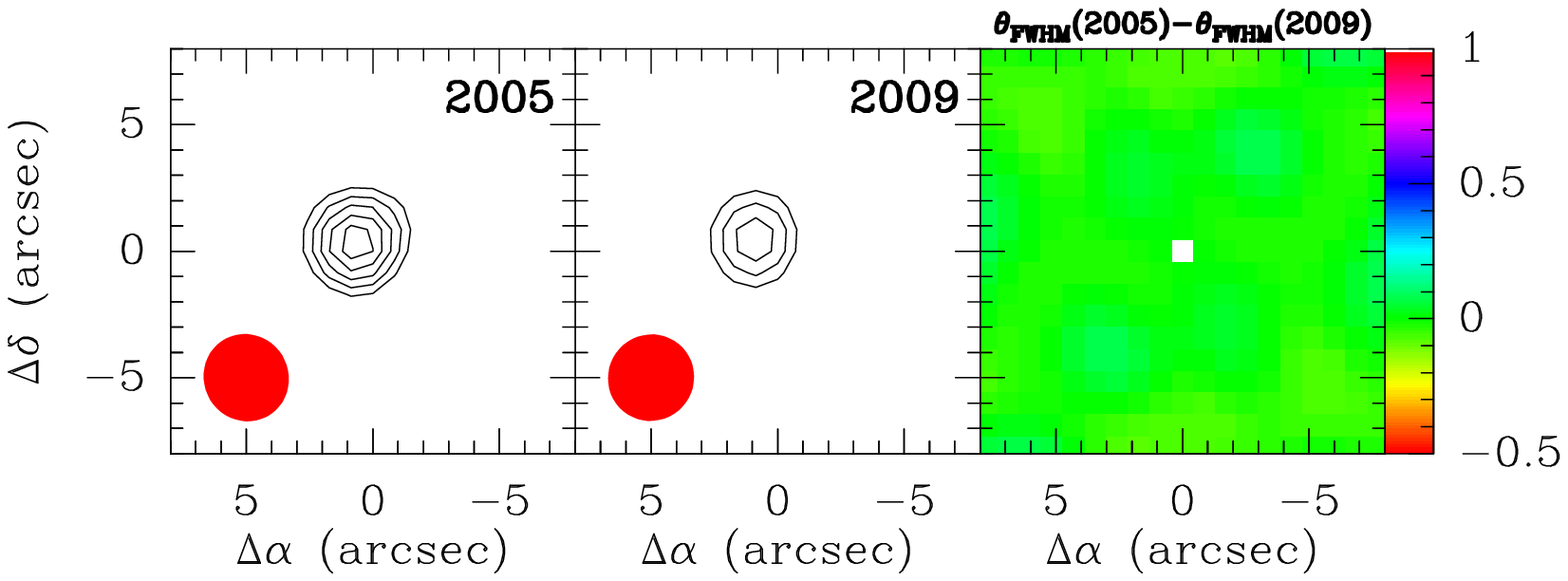}
\caption{Images of the continuum emission at 87$\,$GHz observed with the PdBI toward L1448-mm in the 2005 (left panel) 
and 2009 (middle panel) epochs. The first contour and step levels are 6 (3$\sigma$) and 2$\,$mJy/beam in the 2005 image, and 4.5 (3$\sigma$) and 1.5$\,$mJy/beam in the 2009 map. The peak intensity of the continuum emission measured in 2005 was $\sim$15$\,$mJy/beam, while in 2009 this intensity was $\sim$9$\,$mJy/beam. The synthesized beams are shown in the lower left corner of each panel and are 3.48$"$$\times$3.33$"$, P.A.= 28$^\circ$ for the 2005 epoch, and 3.44$"$$\times$3.37$"$, P.A.=129$^\circ$ for the 2009 map. In the right panel of this Figure, we also show the subtraction of the 2009 beam from the 2005 beam. Color scale and contour levels are as in Figure$\,$\ref{f2}.}
\label{f3}
\end{center}
\end{figure*}

From Figure$\,$\ref{f1} and Table$\,$\ref{tab2}, we find that the SiO line profiles have experienced significant flux variations since 2005 in the 20-50$\,$km$\,$s$^{-1}$ velocity 
interval. Indeed, the measured SiO flux is factors 1.4-3 larger in 2009 for condensations A, B, C, D and E. The absolute differences in the measured SiO flux range from 0.25$\,$Jy$\,$$\,$km$\,$s$^{-1}$ toward E to $\sim$0.8$\,$Jy$\,$$\,$km$\,$s$^{-1}$ toward A, i.e. factors of $\sim$2-7 larger than the estimated error of $\pm$0.12$\,$$\,$Jy$\,$$\,$km$\,$s$^{-1}$. Although falling outside the primary beam of our observations, the averaged SiO line profiles toward E show an evolutionary trend for SiO to show slightly brighter emission at higher velocities in 2009 than in 2005.

One may think that the increase in the SiO flux measured in 2009 with respect to 2005 could be produced by an effect of differences in the data calibration process between both epochs. To check for this potential problem, in Figure$\,$\ref{f3} we compare the images of the continuum emission at 87$\,$GHz obtained in the 2005 and 2009 observations. If calibration effects were responsible for the SiO flux increase, they should be apparent in the continuum images since the continuum emission flux is expected to remain constant with time. From Figure$\,$\ref{f3}, we find that the peak continuum flux for the 2009 epoch is weaker ($\sim$9$\,$mJy/beam) than in the 2005 observations ($\,$15mJy/beam). These peak continuum fluxes are consistent with those previously measured by \citet{gui92} also with the PdBI. Like in the SiO line images, the differences between the 2005 and 2009 beams (right panel in Figure$\,$\ref{f3}) remain below ($\leq$1.4$\,$mJy/beam) the rms noise level in the images (2$\,$mJy/beam in 2005 and 1.5$\,$mJy/beam in 2009). From all this, it is clear that the trend for the SiO line flux to increase from 2005 to 2009 is the opposite to that found for the continuum emission between both epochs, and therefore cannot be attributed to calibration effects. We note that the difference between the continuum peak fluxes measured in 2005 and 2009 ($\sim$6$\,$mJy/beam) is significantly above the rms noise level of the 2005 image, but still within 3$\sigma$.

\begin{figure*}
\begin{center}
\includegraphics[angle=270,scale=0.75]{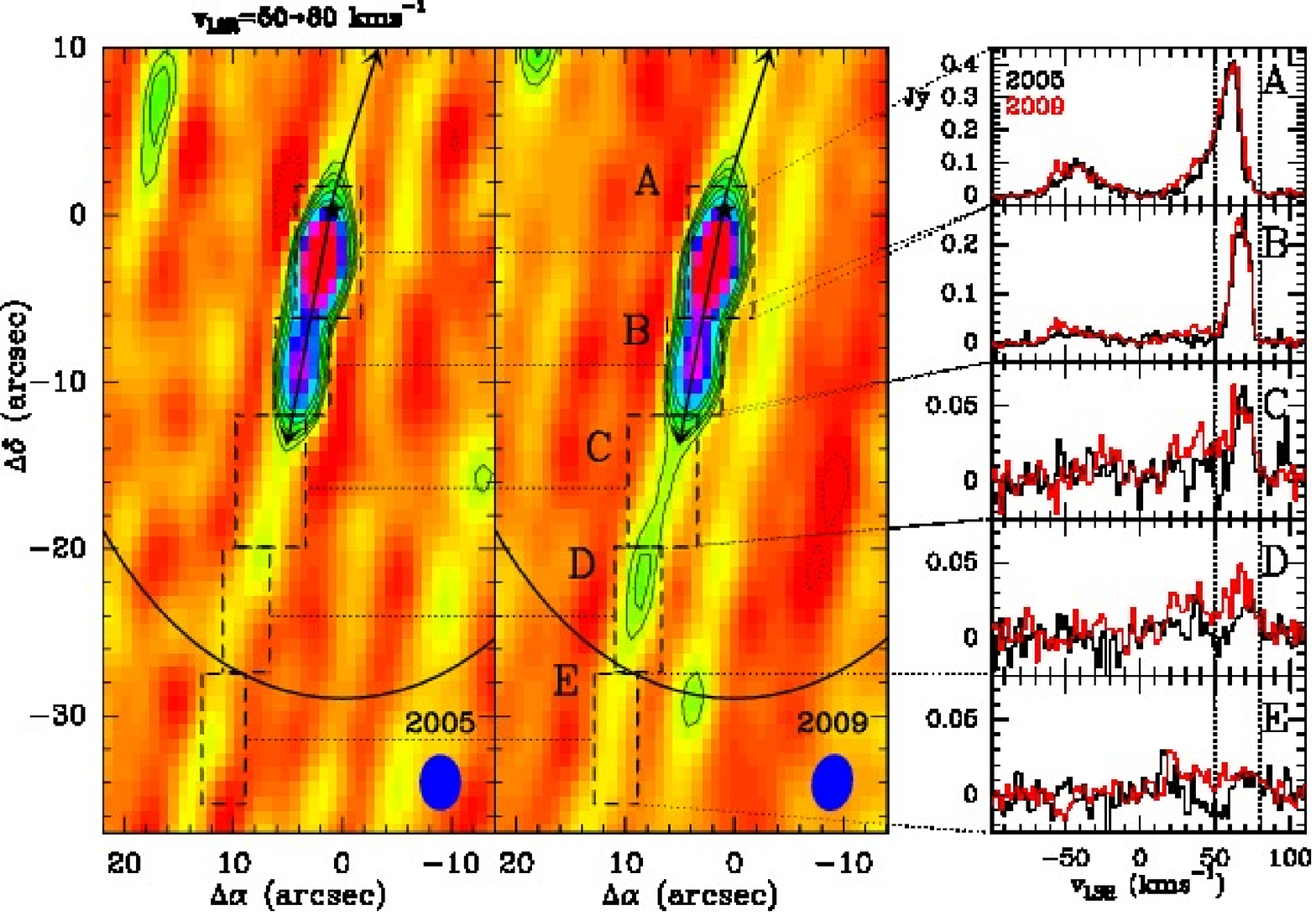}
\caption{As in Figure$\,$\ref{f1}, but for the velocity range from 50 to 80$\,$km$\,$s$^{-1}$ (see vertical dotted lines in the right panels of this Figure). Contour levels in both maps are 13.2 (3$\sigma$), 17.6, 22, 30.8 and 39.6$\,$mJy/beam, with 1$\sigma$=4.4$\,$mJy/beam (Section$\,$\ref{res} and Table$\,$\ref{tab1}). Negative contours correspond to the $-$3$\sigma$ noise level (dotted contours).}
\label{f4}
\end{center}
\end{figure*}

\begin{figure*}
\begin{center}
\includegraphics[angle=270,scale=1.2]{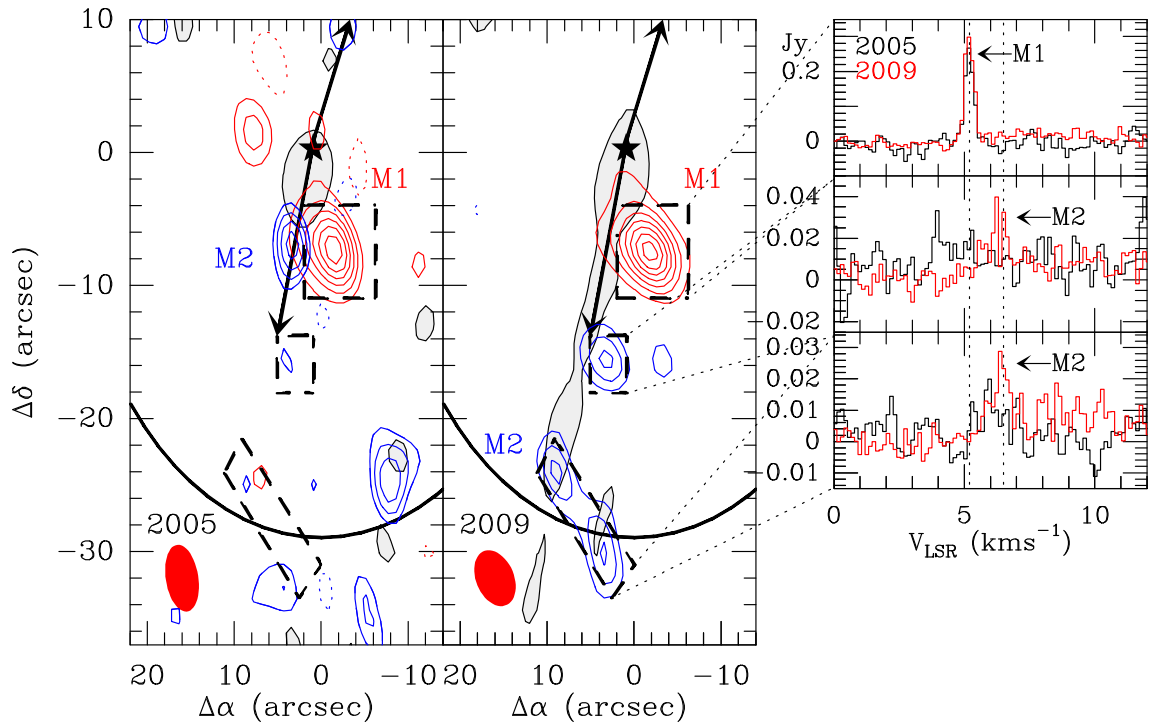}
\caption{Averaged intensity maps of the SiO 
$J$=2$\rightarrow$1 emission observed toward L1448-mm in 2005 (left panel) 
and 2009 (central panel) for the magnetic precursor components at 
5.2$\,$km$\,$s$^{-1}$ (M1; red contours) and at 6.5$\,$km$\,$s$^{-1}$ 
(M2; blue contours). These images are superimposed on the maps
of the intermediate-velocity SiO emission shown in Figure$\,$\ref{f1} (grey
scale and black contours). The first contour and step levels for the M1 component are 23.1 (3$\sigma$) and 23.1$\,$mJy/beam in the 2005 and 2009 maps. The 1$\sigma$ level (7.7$\,$mJy/beam) has been calculated from the thermal noise of the 2005 data (Table$\,$\ref{tab1}).
For the M2 component, the first contour and step levels for the 2005 
image are 17.8 (2$\sigma$) and 8.9$\,$mJy/beam, and for 
the 2009 map, 15.2 (2$\sigma$) and 7.6$\,$mJy/beam. 
The individual rms for every map has been estimated from the thermal
noise of Table$\,$\ref{tab1}.
Filled star and arrows are as in Figure$\,$\ref{f1}. Beams are shown at the lower left corners. Thick ellipse delineate the primary beam of the PdBI observations. Dashed boxes show the regions where the SiO spectra have been averaged (left panels). Dotted vertical lines in the spectra indicate the central radial velocities of the M1 and M2 precursor components.} 
\label{f5}
\end{center}
\end{figure*}

\begin{figure*}
\begin{center}
\includegraphics[angle=270,scale=1.2]{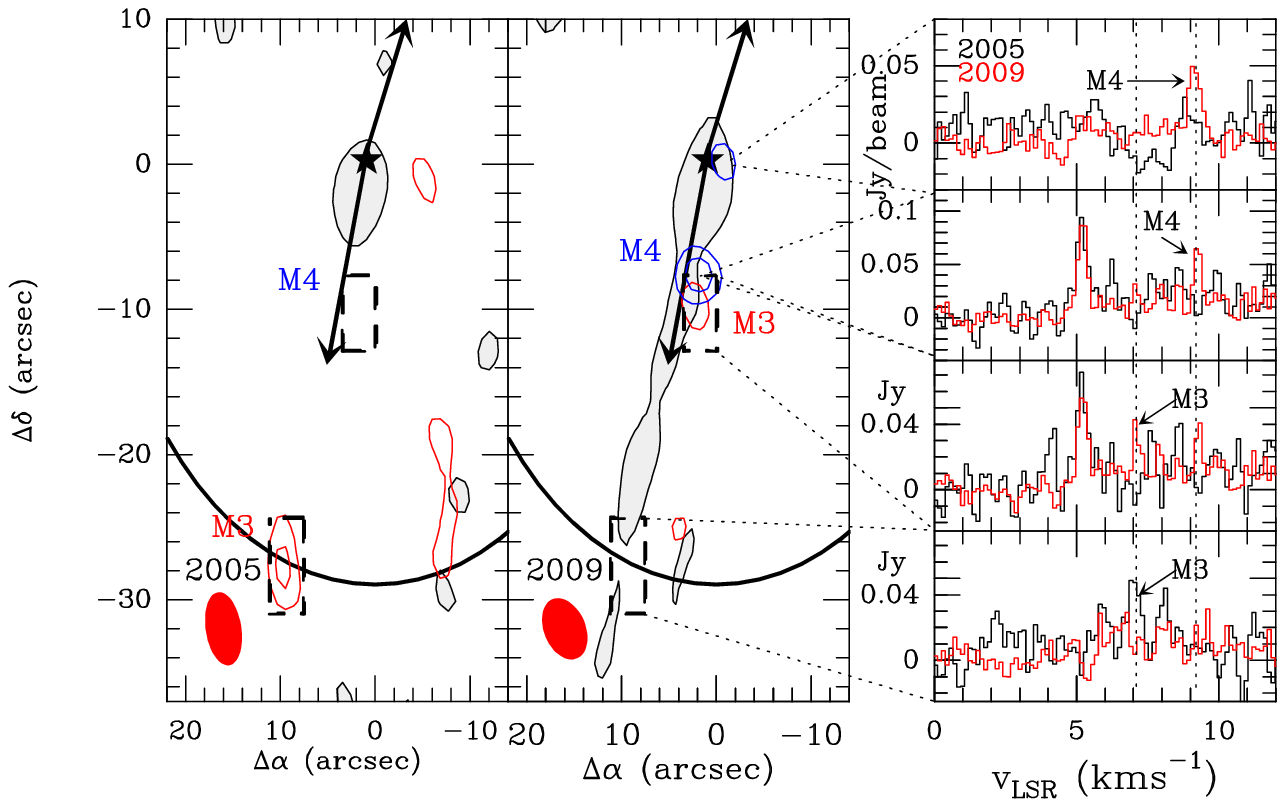}
\caption{Same as in Figure$\,$\ref{f5}, but for the
precursor components M3, at 7.1$\,$km$\,$s$^{-1}$ (red contours), 
and M4, at 9.2$\,$km$\,$s$^{-1}$ (blue contours). The first contour 
and step levels for the M3 component are 25.2 (2$\sigma$) and 
12.6$\,$mJy/beam in the 2005 and 2009 maps. 
For the M4 component, the first contour and step levels are 30.8 (2$\sigma$) 
and 15.4$\,$mJy/beam in the 2005 image, and 26.4 (2$\sigma$) 
and 13.2$\,$mJy/beam for the 2009 map. 
Dashed boxes limit the region where we have averaged the SiO emission 
toward the M3 condensation. For the M4 component, we only show the 
SiO spectra extracted from the emission peaks of the M4 condensations 
(right panels). Dotted vertical lines in the SiO spectra indicate the radial central velocities for M3 and M4.}
\label{f6}
\end{center}
\end{figure*}

From Table$\,$\ref{tab2}, we also find that the absolute differences for 
the measured SiO flux between 50 and 80$\,$km$\,$s$^{-1}$ are similar to those for the velocity range from 30 to 50$\,$km$\,$s$^{-1}$. Indeed, small deviations in the SiO line profiles can also be seen for the 50-80$\,$km$\,$s$^{-1}$ velocity interval (Figure$\,$\ref{f1}). As shown in Figure$\,$\ref{f4}, these deviations do not translate into significant variations in the morphology of the SiO emission toward clumps A and B.
Toward C and D, slightly brighter SiO emission (at the 2-3$\sigma$ intensity level) is seen arising from these condensations. However, as mentioned before,
the variable nature of this emission toward D is unclear, since the derived differences between the 2005 and 2009 beams are larger than 20$\,$mJy/beam (i.e. than the 4$\sigma$ level in the 2005 and 2009 images). 

We note that the SiO flux changes in the SiO emission for the intermediate- and high-velocity shock regimes toward condensations A, B and C, and probably toward E, are likely to be real because very small deviations are found for the low-velocity regime from -10 to 20$\,$km$\,$s$^{-1}$. If these variations were produced by a low-level contribution due to small differences in the UV coverage between the 2005 and 2009 observations, they would appear more dramatically at all velocities which clearly contrasts with our results shown in Table$\,$\ref{tab2}.

\subsection{Variability of the narrow SiO emission at the magnetic 
precursor components}
\label{pre}

Figures$\,$\ref{f5} and \ref{f6} show the averaged intensity images of the SiO 
$J$=2$\rightarrow$1 emission observed in the two epochs
for four different narrow SiO velocity components detected between 
4.7$\,$km$\,$s$^{-1}$ (the ambient cloud velocity) and 
10$\,$km$\,$s$^{-1}$. These components, M1, M2, M3 and M4, have very narrow line 
profiles (linewidths of $\sim$0.25-0.7$\,$km$\,$s$^{-1}$), and 
are centered at radial velocities 5.2, 6.5, 7.1 and 
9.2$\,$km$\,$s$^{-1}$, respectively. The rms noise level in every image
has been calculated from the thermal noise
shown in Table$\,$\ref{tab1} divided by $\sqrt{N}$ with N the number of 
channels used to generate the averaged intensity map. For the M1 and M3 components, we consider as reference the rms noise measured in the 2005 observations (i.e. 7.7$\,$mJy/beam for M1 and 
12.6$\,$mJy/beam for M3). For the M2 and M4 components, in general fainter than the M1 and M3 narrow emission, we have used the individual rms noise level obtained in every image (for M2, 8.9$\,$mJy/beam and 7.6$\,$mJy/beam in the 2005 and 2009 images, respectively; and for M4, 15.4$\,$mJy/beam and 13.2$\,$mJy/beam in the 2005 and 2009 maps).

From Figure$\,$\ref{f5}, we find that the narrow SiO emission at 
5.2$\,$km$\,$s$^{-1}$ (i.e. the M1 component) with a linewidth
of only $\sim$0.4$\,$km$\,$s$^{-1}$, is the only component clearly 
detected in both epochs. This emission, previously reported by \citet{jim04} from single-dish IRAM 30$\,$m observations, was interpreted as 
the signature of the magnetic precursor of young C-shocks toward 
the L1448-mm outflow. The M1 emission peaks at (-2,-7), and although compact, 
shows an elongation in the direction perpendicular to the 
high-velocity jet (size of $\sim$4$''$$\times$6$''$, 
0.005$\,$pc$\times$0.007$\,$pc). We note that this emission falls to the north of the secondary CO $J$=3$\rightarrow$2 outflow reported by \citet{hir10} and powered by the south-eastern source L1448-C(S) \citep{jor06}. In addition, the narrow SiO kinematics do not exactly match those of the CO $J$=3$\rightarrow$2 emission, suggesting that the M1 emission is not associated with this secondary outflow.

From the averaged SiO spectra of Figure$\,$\ref{f5} (right panels), the derived SiO 
integrated flux for M1 in the 2009 spectrum is slightly brighter than in the 
2005 data by 20\%$\pm$13\%. The error in the relative difference of the 
SiO flux is large. However, as shown in Section$\,$\ref{mod}, the flux increase by 20\% should be considered as a lower limit since the continuum peak intensity in the 2009 images is weaker than in 2005. Therefore, it is possible that the narrow SiO emission arising from M1 is also variable.

In contrast with M1, the other three narrow SiO components (M2, M3 and M4)  
seem to be associated with transient structures that appear either in  
the 2005 or 2009 images. For instance, for the M2
narrow emission at 6.5$\,$km$\,$s$^{-1}$, one condensation is 
found toward clump B in the 2005 map, while two new M2 condensations
appear associated with clumps C and with the south-east edge of condensation 
D. From the SiO spectra averaged over these new condensations 
(Figure$\,$\ref{f5}), we find that the SiO line profiles seem to have
evolved since 2005 because their emission peaks appear more red-shifted 
within the shock, and their line wings are slightly brighter 
at velocities ranging from 7 to 10$\,$km$\,$s$^{-1}$. 
In addition, the M2 condensation toward D is particularly interesting because it also shows an elongation in the direction perpendicular to the jet, as if tracing the interface between the recently shocked gas of clumps D and E. 

We note that the detection of the M2 emission components in 2009 could be a direct consequence of the higher sensitivity of the 2009 images. However, no similar structure is seen in the 2005 map above the 2$\sigma$ level 
(17.8$\,$mJy/beam; left panel of Figure$\,$\ref{f5}), 
which is below the 
3$\sigma$ noise level (of 22.8$\,$mJy) of the 2009 image. 
This suggests that the M2 narrow SiO lines, with linewidths of 
0.7$\,$km$\,$s$^{-1}$, were likely not present in 2005.   

The M3 and M4 narrow SiO components (see Figure$\,$\ref{f6}) 
have linewidths of $\sim$0.25-0.6$\,$km$\,$s$^{-1}$ and their
morphology is more compact (sizes of $\leq$4$''$). The M3 and 
M4 clumps are found toward A and B, likely forming part of a 
larger C-shock structure (left panels of Figure$\,$\ref{f6}). 
The M3 emission seen at the interface between clumps D and 
E in 2005 seems to have vanished in the 2009 image, revealing
the transient nature of the precursor phenomenon. 
It is remarkable that the M4 clump close to the L1448-mm 
protostar is relatively bright (peak intensity of $\sim$0.05$\,$Jy/beam) 
but remains unresolved in the 2009 images. This
implies that the magnetic precursor region toward this component 
is smaller than $\sim$3.5$''$ (0.004$\,$pc). 

\begin{deluxetable*}{ccccccccccc}
\tabletypesize{\scriptsize}
\tablecaption{Ratios and absolute differences in the measured SiO flux between the 2005 and 2009 spectra.\label{tab2}}
\tablewidth{0pt}
\tablehead{
& & \multicolumn{3}{c}{$F$ [Jy$\,$km$\,$s$^{-1}$]} & \multicolumn{3}{c}{$F$(2009)/$F$(2005)} & \multicolumn{3}{c}{$|F$(2009)-$F$(2005)$|$ [Jy$\,$km$\,$s$^{-1}$]}\\ \cline{3-5} \cline{6-8} \cline{9-11}
 & Epoch & \colhead{-10$\rightarrow$20} & \colhead{20$\rightarrow$50} & \colhead{50$\rightarrow$80} & \colhead{-10$\rightarrow$20} & \colhead{20$\rightarrow$50} & \colhead{50$\rightarrow$80} & \colhead{-10$\rightarrow$20} & \colhead{20$\rightarrow$50} & \colhead{50$\rightarrow$80} \\
& & (km$\,$s$^{-1}$) & (km$\,$s$^{-1}$) & (km$\,$s$^{-1}$) & (km$\,$s$^{-1}$) & (km$\,$s$^{-1}$) & (km$\,$s$^{-1}$) & (km$\,$s$^{-1}$) & (km$\,$s$^{-1}$) & (km$\,$s$^{-1}$)}
\startdata
{\bf A} & 2005 & $\leq$0.2\tablenotemark{a} & 1.93(7)\tablenotemark{b} & 5.97(7) & $\dots$ & 1.40$\pm$0.05 & 1.106$\pm$0.015 & $\dots$ &  0.76$\pm$0.12 & 0.63$\pm$0.12 \\
        & 2009 & $\leq$0.15 & 2.69(5) & 6.60(5) & & & \\
{\bf B} & 2005 & 0.24(7) & $\leq$0.2 & 3.33(7) & $\leq$0.63 & $\geq$3.3$\pm$1.1 & 1.09$\pm$0.03 & $\leq$0.09 & 0.45$\pm$0.12 & 0.29$\pm$0.12 \\
        & 2009 & $\leq$0.15 & 0.65(5) & 3.63(5) & & \\
{\bf C} & 2005 & $\leq$0.2 & 0.23(7) & 0.63(7) & $\dots$ & 3.1$\pm$0.9 & 1.6$\pm$0.2 & $\dots$ & 0.47$\pm$0.12 & 0.37$\pm$0.12 \\
        & 2009 & $\leq$0.15 & 0.71(5) & 1.00(5) & & & \\
{\bf D} & 2005 & $\leq$0.2 & 0.25(7) & 0.31(7) & $\geq$0.8 & 2.7$\pm$0.7 & 2.6$\pm$0.6 & $\leq$0.04 & 0.43$\pm$0.12 & 0.49$\pm$0.12 \\
        & 2009 & 0.16(5) & 0.68(5) & 0.80(5) & & \\
{\bf E} & 2005 & $\leq$0.2 & $\leq$0.2 & $\leq$0.2 & $\dots$ & $\geq$2.3$\pm$0.8 & $\geq$1.9$\pm$0.7 & $\dots$ & 0.25$\pm$0.12 & 0.18$\pm$0.12 \\
        & 2009 & $\leq$0.15 & 0.44(5) & 0.37(5) & &
\enddata
\tablenotetext{a}{Upper limits correspond to the 3$\sigma_{Area}$ level, where $\sigma_{Area}$ is calculated as $\sigma$$\sqrt{\delta v \Delta v}$ (see Section$\,$\ref{mod}).}
\tablenotetext{b}{Uncertainties in the SiO integrated intensity flux correspond to the $\sigma_{Area}$ level.}
\end{deluxetable*}

\section{Origin of the variability of the thermal SiO line emission toward 
L1448-mm}
\label{origin}

Variability of the atomic/molecular emission toward 
molecular outflows have usually focused on the detection of proper motions 
associated with the high-velocity shocked gas in these objects 
\citep[][]{rei01,gir01,mcg07}. However, it has usually been neglected that 
the moderate- and intermediate-velocity shocked gas generated in C-shocks, could also be subject to changes in the line emission arising from these regions. 

Our data show that the SiO emission at intermediate velocities toward 
the red-shifted lobe of the L1448-mm outflow has changed in morphology, line intensity, and line shape in only 4$\,$yrs. This is particularly true
for condensations A, B and C, where their measured SiO flux increase by factors 1.4-3 cannot be attributed to systematic differences between the beams of the observations 
(Section$\,$\ref{mod}). This flux increase suggests that new material has likely entered the shock, enhancing the gas-phase abundance of SiO toward these regions. 
The lack of detection of SiO in previous interferometric observations toward the location of the new SiO condensations \citep{dut97}, also favours the idea of a recent SiO enhancement toward the red-shifted lobe of the L1448-mm outflow. Although not predicted by the models of \citet{jim09} for such short time-scales, the SiO line profiles could also be variable at high velocities with a moderate increase in their integrated intensity flux.

The origin for such a coherent variability in the SiO emission over spatial
scales of 750-1000$\,$AU (i.e. over the 30-40$"$ mapped with the PdBI toward the red-shifted lobe of the L1448-mm outflow) could be related to the formation of C-shocks in the internal working surface between the high-velocity jet and the ambient material. These shocks would tend to propagate sideways and would participate in the expansion processes associated with the high-velocity jet 
\citep[see e.g.][]{cab95}. As shown by the extended SiO and H$_2$ shock-excited emission reported by \citet{dut97} and \citet[][]{neu09}, the L1448-mm jet has already travelled along distances larger than $\geq$2$'$, explaining the coverage of at 
least 30-40$"$ where variability of the SiO emission has been detected. In Section$\,$\ref{mod}, we have shown that the deconvolved size of the B, C, D and E condensations in the direction perpendicular to the high-velocity jet is $\leq$1.8$"$ (i.e. 0.002$\,$pc at a distance of 250$\,$pc). Assuming that C-shocks generated at the working surface travel symmetricaly and sideways half that distance from the jet,
we estimate a propagation velocity for those C-shocks of 
$\leq$240$\,$km$\,$s$^{-1}$ in the 4$\,$yrs time-interval between the 2005 and 2009 observations. These velocities are consistent with those derived by \citet[][of few hundred km$\,$s$^{-1}$]{gir01} from the proper motions of the SiO high-velocity bullets observed toward the red-shifted lobe of the L1448-mm outflow.

We would like to stress that the presence of C-shocks with such high velocities could be possible within an scenario of a highly magnetized medium with a large pre-shock magnetic induction, $B$, transverse to the direction of the flow. Indeed, \citet{leb02} showed that, for large values of $B$ (of the order of some mG), the critical velocities of C-shocks can be as high as few 100$\,$km$\,$s$^{-1}$, which are similar to those derived from our observations.

The variability of the thermal SiO emission toward L1448-mm is also supported by our high-spectral resolution SiO data, which show regions (the M2 precursor component) where the narrow SiO line profiles seem to have evolved since 2005. The emission peak of these lines have been red-shifted within the shock, and have developed slightly brighter broad SiO emission as predicted by shock models \citep{jim09}. Although the broad SiO condensation E falls outside the primary beam of our observations, this 
trend in the evolution of the SiO line profile since 2005 is also found
toward this condensation.


The narrow SiO emission toward the M1 component also seems to have experienced
a small increase in its integrated intensity flux (by 20$\pm$13\%). This component, previously reported by single-dish observations \citep{jim04}, is known to be 
correlated with an enhancement and over-excitation of the ion fluid (represented by 
the H$^{13}$CO$^{+}$ molecule) with respect to the neutral one \citep[with molecules such as HN$^{13}$C and H$^{13}$CN; see e.g.][]{jim06,rob10}, as expected within the magnetic precursor scenario. However, although predicted by our C-shock modelling results \citep{jim09}, no significant variation in the linewidth of narrow SiO is detected 
for the M1 component toward the L1448-mm outflow.


The other narrow SiO components (M2, M3 and M4) are fainter and their linewidths are as narrow as $\sim$0.25$\,$km$\,$s$^{-1}$. 
Even in the case that no turbulence was present in the medium, these
linewidths would imply a kinetic temperature of $\leq$60$\,$K 
for the SiO gas, which could only be attained at   
the early precursor stage or at the evolved postshock gas 
\citep[see e.g.][]{flo03,gus08}. Since these components seem
to be transient (Section$\,$\ref{pre}), it is likely that they arise from the precursor region of the shock. Reflected shock waves within the postshock gas could not be responsible 
for the narrow SiO lines in this outflow, because their ion-neutral 
velocity drift never exceeds 6$\,$km$\,$s$^{-1}$ \citep[][]{ash10}. This is
the lower limit drift velocity required to sputter detectable 
abundances of SiO in the gas-phase, from the mantles of dust grains
\citep[][]{cas97,jim08,jim09}. 

The morphology of the M2 emitting regions appears elongated in the direction perpendicular to the high-velocity jet, resembling the bow-shock structures 
observed toward other jets and outflows \citep[e.g. as in HH1-2 or Orion;][]{eis94,dav09}. In addition, these regions remain practically unresolved 
in the direction of the jet ($\leq$4$''$, 1.5-1.8$\times$10$^{16}$$\,$cm;
Section$\,$\ref{pre}), and are generally found at the interface between the recently shocked condensations A, B, C, D and E detected at intermediate velocities in the 2009 maps. This indicates that the interface/precursor 
regions typically have very small length scales as predicted by C-shock modeling 
\citep[$\sim$3-20$\times$10$^{15}$$\,$cm;][]{gus08,jim08}.
This is also consistent with the fact that the M4 narrow SiO line arises
from unresolved regions (Section$\,$\ref{pre}). 

We note that, although we compare our results with those from 
steady-state C-shock models, recent modeling of the transient evolution of C-shocks 
predict the presence of a magnetic precursor within the 
physical structure of the shock, as a consequence of its propagation through a 
decreasing density medium such as that found in molecular outflows \citep[][]{ash10}. 

\section{Conclusions}
\label{con}

In this paper, we present interferometric images of the thermal emission of 
the SiO $J$=2$\rightarrow$1 rotational line observed toward the 
L1448-mm outflow at two different epochs (2005 and 2009). The PdBI images reveal
that this emission is variable in only 4$\,$yrs. 
Not only the morphology of the SiO emission, but also its line profiles, have 
significantly changed since 2005, with a clear increase of the SiO flux
at intermediate-velocities, and possibly at high-velocities. This suggests 
that new material has recently entered the shock. 

For the precursor component, we have detected several narrow SiO velocity 
components (M1, M2, M3 and M4) with an elongated morphology
in the direction perpendicular to the high-velocity jet. These features 
resemble those found in bow-shocks. Although these components could have  
appeared recently, new multi-epoch and higher sensitivity interferometric images 
are needed to clearly establish the variable nature of the narrow SiO emission 
detected toward L1448-mm. 

We speculate that the variability of the thermal SiO $J$=2$\rightarrow$1 emission observed toward the L1448-mm outflow could be explained by the 
time-dependent propagation of very young C-shocks 
(with their precursors) through the ambient molecular cloud. Monitoring 
of the variability of the thermal SiO emission toward young molecular outflows
could be a useful tool to study the evolution of these objects, and their
impact on the surrounding medium.

\acknowledgments

We would like to thank the IRAM director, P. Cox, for letting us
carry out the second epoch of the PdBI SiO observations, 
despite the reticence of the IRAM Programme Committee. 
We acknowledge an anonymous referee for his/her critical 
reading and constructive comments that 
helped to largely improve the manuscript.
IJ-S acknowledges the Smithsonian Astrophysical Observatory for the
support provided through a SMA fellowship. JM-P and IJ-S have been 
partially funded by MICINN grants ESP2007-65812-C02-C01 and 
AYA2010-21697-C05-01 and AstroMadrid (CAM S2009/ESP-1496).


\end{document}